# On Beat Frequency Oscillation of Two-Stage Wireless Power Receivers

Kerui Li, *Student Member, IEEE*, Siew-Chong Tan, *Senior Member, IEEE,* and Ron Shu Yuen Hui, *Fellow, IEEE*

***Abstract*-Two-stage wireless power receivers, which typically include an AC-DC diode rectifier and a DC-DC regulator, are popular solutions in low-power wireless power transfer applications. However, the interaction between the rectifier and the regulator may introduce beat frequency oscillation on both the DC-link and output capacitors. In this paper, the cause of the beat frequency oscillation and its related issues are investigated with the corresponding design solution on alleviating the oscillation discussed. Theoretical and experimental results verifying the presence of beat frequency oscillation in the two-stage wireless receiver system are provided. Our study shows that the beat frequency oscillation can be significantly alleviated if appropriate design solutions are applied.**

## I. Introduction

The two-stage wireless power receiver, comprising a passive diode rectifier and a DC-DC converter, is currently the most popular solution for low-power consumer electronic products with wireless power transfer (WPT) capability [1]-[3]. The first-stage passive diode rectifier converts the AC current into DC current, of which the DC-link capacitor buffers the ripples to provide a relatively constant DC voltage to the second stage. The second-stage DC-DC converter is controlled to provide a constant current or voltage to charge the battery load.

Despite this solution being conceptually simple and effective, the potential beat frequency voltage fluctuation caused by the interaction between the diode rectifier and the post regulated DC-DC converter may adversely affect its performance. In practice, the operating frequency of the passive diode rectifier $f_1$ and the operating frequency of the DC-DC converters $f_2$ are not necessarily identical. The beat frequency ($f_1$−$f_2$ or $f_2$−$f_1$) components between the two can lead to a beat frequency voltage fluctuation on the DC-link capacitor. This fluctuation can be propagated to the output capacitor of the DC-DC converter [4], contaminating the output voltage with a beat frequency oscillation, which may breach the output regulation requirement and even causing damage to the load. Moreover, in high power applications, having high voltage and high current naturally introduce more high-frequency ringing due to the inherent circuit parasitic components, which may further aggravate the output oscillation [5].

The phenomenon of beat frequency oscillation has been investigated [6], modelled [7]-[9] and analysed [10]-[12] in DC-DC converter systems. The frequency difference between DC-DC converters may amplify the side-band harmonics, which therefore leads to substantial voltage/current oscillations. However, these studies have never been reported on the wireless power receiver system. The high-frequency AC input nature of the wireless power receiver system fails to comply with the usual DC input assumptions adopted for modelling and analysing cascaded DC-DC converters. Due to its AC nature, the modelling and analysis of the two-stage wireless power receiver system is more challenging than that of those in cascaded DC-DC converter systems. This is because the first-stage passive diode bridge rectifier can only provide varied and discontinuous current to the DC-link capacitor, which leads to significant voltage ripples containing rich frequency components. Subsequently, the modelling and analysis of the wireless power receiver system involves not only the DC component, but the beat frequency and switching-frequency components.

In this work, the phenomenon of the beat frequency oscillation in the two-stage wireless power receiver is investigated. The time-domain and multi-frequency models of the two-stage wireless power receiver system, as well as the analysis of the beat frequency oscillation, are provided. In addition, possible solutions on alleviating the oscillation in the two-stage wireless power receiver system, are discussed.

## II. Model of The Two-stage Wireless Power Receiver System

### *A. Two-Stage Wireless Power Receiver System*

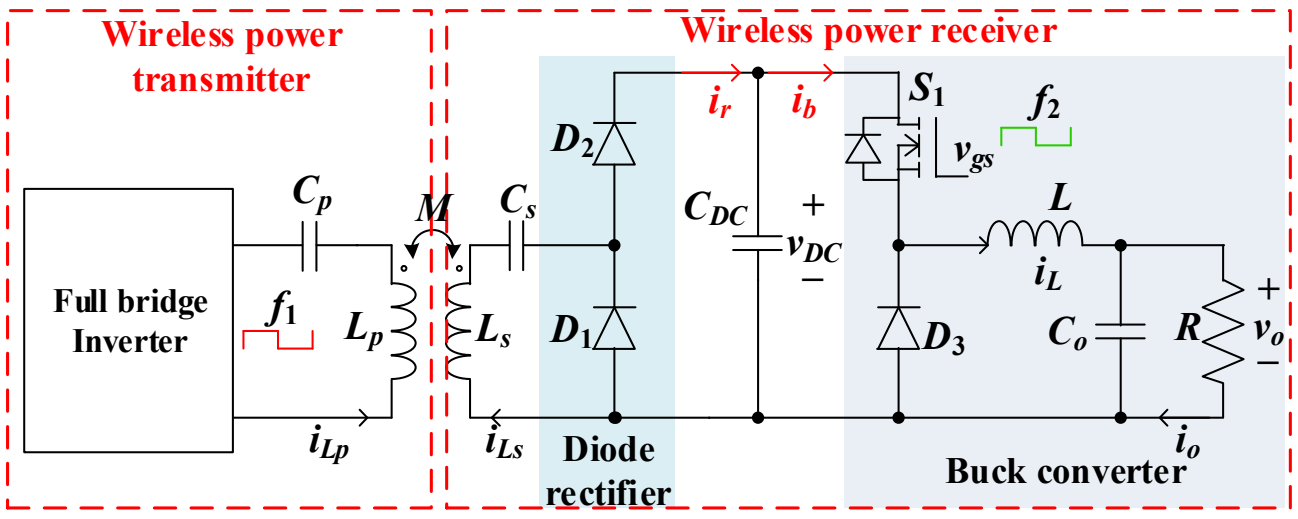

Fig. 1. Schematic diagram of a WPT system with the two-stage wireless receiver.

Fig. 1 shows the circuit configuration of a WPT system with the two-stage wireless receiver. The wireless power transmitter comprises a full-bridge inverter (which generates a square AC voltage waveform with a frequency of $f_1$) and a primary-side transmitter coil $L_p$ with a series compensation capacitor $C_p$ (with resonant frequency being equal to the switching frequency, i.e., $1/\sqrt{C_pL_p} = 2\pi f_1$ [13]). The wireless power receiver contains the secondary-side receiver coil $L_s$ with a series compensation capacitor $C_s$ (designed to resonate at $f_1$, i.e. $1/\sqrt{C_sL_s} = 2\pi f_1$ [13]), a first-stage diode rectifier ($D_1$ *and* $D_2$) for converting AC power into an unregulated DC voltage $v_{DC}$ at the DC-link capacitor $C_{DC}$, and a second-stage buck converter (comprising switch $S_1$, diode $D_3$, output inductor $L$, output capacitor $C_o$, and load $R$). The buck converter operates independently at switching frequency $f_2$ and duty cycle $D$ to output a regulated voltage $v_o$ to the load. In practice, the wireless transmission frequency and the buck converter frequency are typically different, i.e., $f_1 \neq f_2$. Attributed to this frequency difference, a beat frequency $f_b$ ( $|f_1-f_2|$ ) oscillation is introduced into the DC-link capacitor and output capacitor.

### *B. Operating Principle and Time-Domain Model*

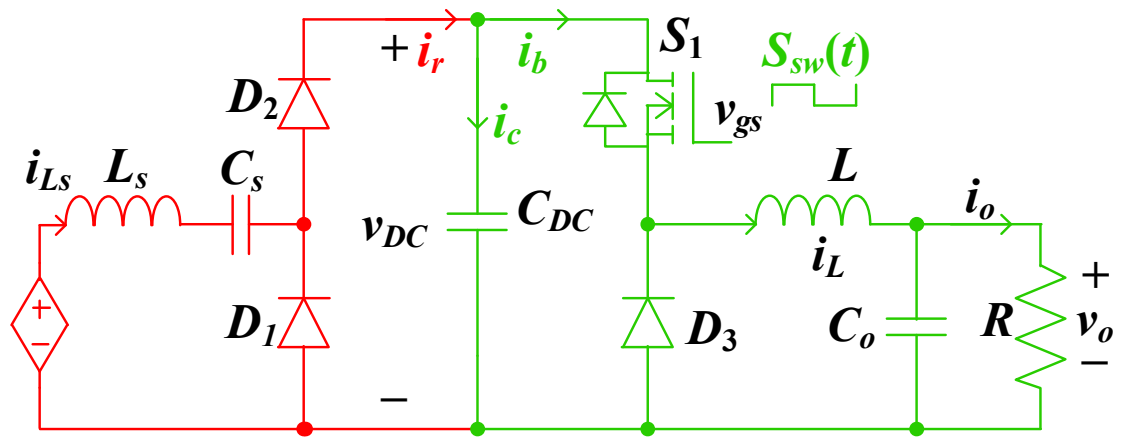

Fig. 2. Equivalent circuit model of the two-stage wireless power receiver.

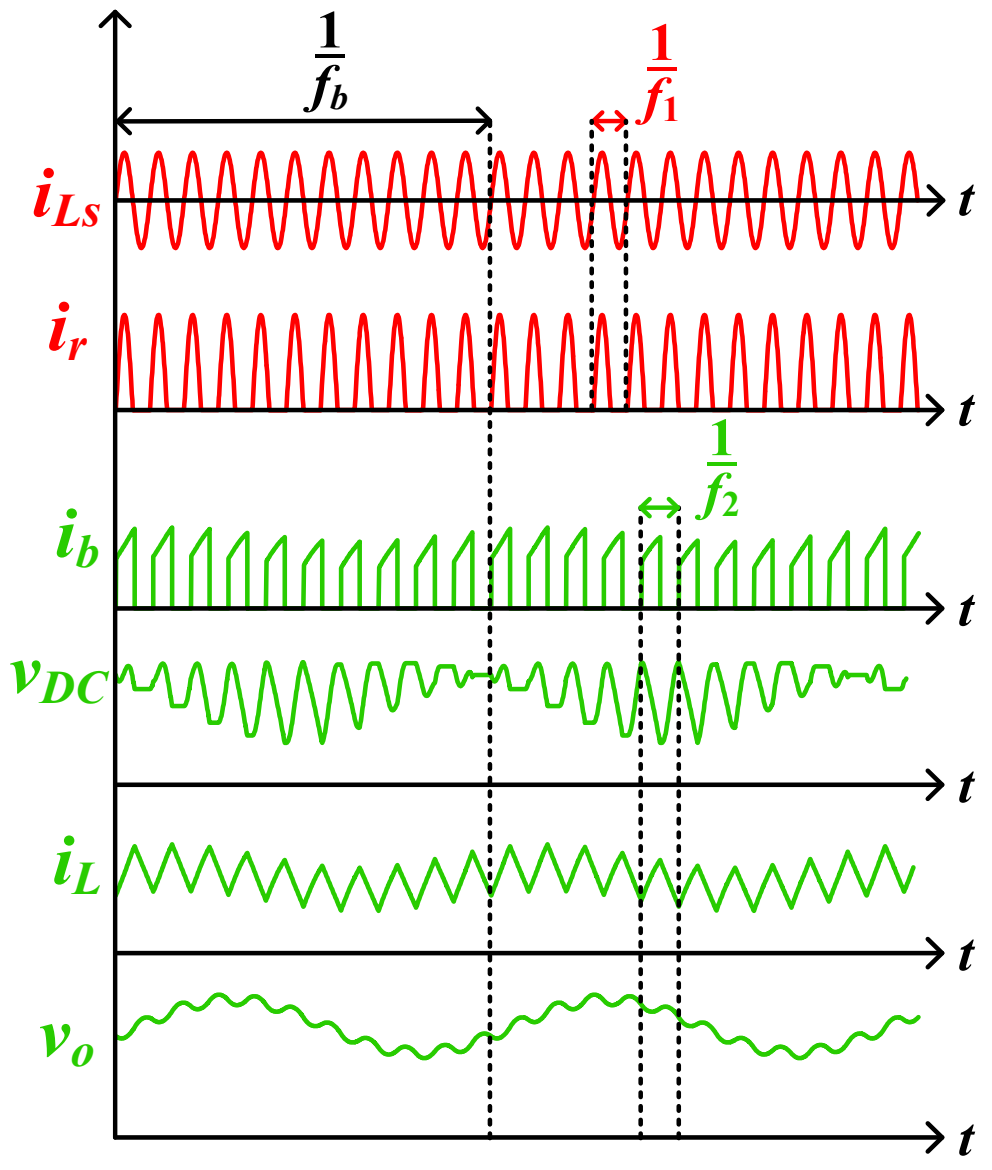

Fig. 3. Key waveforms of the two-stage wireless power receiver (where $1.1f_2 = f_1$) with the beat frequency oscillation.

Fig. 2 shows the equivalent circuit model of the two-stage wireless power receiver. Fig. 3 shows the key waveforms of the system with beat frequency oscillation in the case where $1.1f_2 = f_1$. Using series-series compensation, the wireless power receiver coil $L_s$ introduces a sinusoidal current $i_{Ls}(t)$ with the frequency of $f_1$ to the diode bridge rectifier, i.e.,

$$i_{Ls}(t) = I_{Ls}\sin(2\pi f_1 t), \tag{1}$$

where $I_{Ls}$ is the amplitude of the sinusoidal current $i_{Ls}(t)$.

Here, the diode rectifier has two operating modes. When $i_{Ls}(t)>0$, the current flows to the DC-link capacitor $C_{DC}$ via diode $D_2$. When $i_{Ls}(t)<0$, the current flows through the freewheeling diode $D_1$, disconnecting the inductor from $C_{DC}$. Subsequently, the rectified current $i_r(t)$ may be expressed as

$$i_r(t) = \begin{cases} I_{Ls}\, sin(2\pi f_1 t), & if\ \frac{n}{f_1} < t \le \frac{n+0.5}{f_1}, n \in \mathbb{Z} \\ 0, & if\ \frac{n+0.5}{f_1} < t \le \frac{n+1}{f_1}, n \in \mathbb{Z} \end{cases} \tag{2}$$

Following the passive rectifier stage is the buck converter, which operates at switching frequency $f_2$. Its switching signal $s_{sw}(t)$ is given as

$$s_{sw}(t) = \begin{cases} 1, & if\ \frac{n}{f_2} < t \le \frac{n+D}{f_2}, n \in \mathbb{Z} \\ 0, & if\ \frac{n+D}{f_2} < t \le \frac{n+1}{f_2}, n \in \mathbb{Z} \end{cases} \tag{3}$$

The DC-link capacitor $C_{DC}$ buffers the difference between the rectified current $i_r(t)$ of the passive diode rectifier and the discontinuous current $i_b(t)$ of the buck converter, as well as maintains a relatively constant DC voltage $v_{DC}(t)$. The differential equation of the DC-link capacitor voltage $v_{DC}(t)$ is expressed as

$$C_{DC}\frac{\partial v_{DC}(t)}{\partial t} = i_r(t) - i_b(t) = i_r(t) - s_{sw}(t)i_L(t) \tag{4}$$

Attributing to the beat frequency between the passive rectifier and buck converter, a beat frequency ($f_b$) oscillation is observed on $v_c(t)$, as shown in Fig. 3.

The inductor current $i_L(t)$ of the buck converter is governed by the differential equation

$$L\frac{\partial i_L(t)}{\partial t} = s_{sw}(t)v_{DC}(t) - v_o(t) \tag{5}$$

The switching function $s_w(t)$ modulates $v_{DC}(t)$ and propagates its DC component, switching frequency component at $f_2$, and the beat frequency component at $f_b$, to the inductor current $i_L(t)$. Consequently, as shown in Fig. 3, $i_L(t)$ contains not only the DC and switching frequency component at $f_2$, but also the beat frequency current fluctuation.

The output capacitor $C_o$ buffers the switching ripples (at the frequency of $f_2$) of $i_L(t)$ and bypass the DC and beat frequency component to the load resistor $R$. The differential equation of the output voltage $v_o(t)$ is written as

$$C_o\frac{\partial v_o(t)}{\partial t} = i_L(t) - \frac{v_o(t)}{R} \tag{6}$$

Eq. (6) reflects that $C_o$ and $R$ behave as an $RC$ low-pass filter. However, the beat frequency components may bypass this filter and propagate to the output voltage. Consequently, $v_o(t)$ contains not only high frequency switching ripples (at $f_1$), but the beat frequency oscillation. However, it is difficult to directly obtain the required multi-frequency ($f_b$, $f_1$, and $f_2$) components from the time-domain model (Eq. (1)-(6)).

### *C. Multi-Frequency Model*

The existing modelling methods reported in [14] and [15] lack frequency resolution to holistically observe the multi-frequency (DC, $f_b$, $f_1$, and $f_2$) components. Consequently, to enhance the frequency resolution in multi-frequency modelling, the fundamental frequency $f_{base}$ in the multi-frequency model is defined as the greatest common divisor of $f_1$ and $f_2$, i.e., $f_1=M_1 f_{base}$, and $f_2=M_2 f_{base}$, where $M_1$ and $M_2$ are positive integers. The beat frequency $f_b$ is therefore represented as $f_b=|M_1-M_2|f_{base}$. Subsequently, the multi-frequency components (DC, $f_b$, $f_1$, and $f_2$) are represented by the harmonic components of $f_{base}$.

The conversion from the time-domain model to the multi-frequency model contains three categories: 1. Fourier coefficients decomposition of $i_{Ls}(t)$, and $i_r(t)$, $v_{DC}(t)$, $i_L(t)$ and $v_o(t)$; 2. Multi-frequency-domain convolution of the multiplication terms $S_{sw}(t)i_L(t)$ and $S_{sw}(t)v_{DC}(t)$; 3. Multi-frequency-domain differentiation of the differentiation terms $C_{DC}\frac{\partial v_{DC}(t)}{\partial t}$, $L\frac{\partial i_L(t)}{\partial t}$, and $C_o\frac{\partial v_o(t)}{\partial t}$.

*1: Fourier coefficients decomposition.* Without loss of generality, $f_1$ is assumed to be greater than $f_2$, i.e., $M_1$>$M_2$ (note that the derivation process and model presented as follows are valid for the cases of $f_1 < f_2$). The beat frequency $f_b$ is therefore represented as $f_b$=($M_1$−$M_2$)$f_{base}$. Taking $f_{base}$ as the fundamental frequency, the Fourier expansion of the time-domain variables of $i_{Ls}(t)$ and $i_r(t)$ can be written as

$$i_{Ls}(t) \approx \sum_{k=-2M_1}^{2M_1} I_{Ls<k>}e^{i2\pi k f_{base}t} = \boldsymbol{E}\boldsymbol{I_{Ls}} \tag{7}$$

$$i_r(t) \approx \sum_{k=-2M_1}^{2M_1} I_{r<k>}e^{i2\pi k f_{base}t} = \boldsymbol{E}\boldsymbol{I_r} \tag{8}$$

of which the Fourier basis $\boldsymbol{E}$ is

$$\boldsymbol{E} = [e^{i4M_1\pi f_{base}t},\dots,e^{i4\pi f_{base}t},e^{i2\pi f_{base}t},1,e^{-i2\pi f_{base}t},e^{-i4\pi f_{base}t},\dots,e^{-i4M_1\pi f_{base}t}] \tag{9}$$

and the corresponding Fourier coefficients $\boldsymbol{I_{Ls}}$ and $\boldsymbol{I_r}$ are

$$\boldsymbol{I_{Ls}} = [I_{Ls_{<2M_1>}},\dots,I_{Ls_{<2>}},I_{Ls_{<1>}},I_{Ls_{<0>}},I_{Ls_{<-1>}},I_{Ls_{<-2>}},\dots,I_{Ls_{<-2M_1>}}]^T \tag{10}$$

$$\boldsymbol{I_r} = [I_{r_{<2M_1>}},\dots,I_{r_{<2>}},I_{r_{<1>}},I_{r_{<0>}},I_{r_{<-1>}},I_{r_{<-2>}},\dots,I_{r_{<-2M_1>}}]^T \tag{11}$$

Since the time-domain expression of $i_{Ls}(t)$ and $i_r(t)$ are explicit, the elements of $\boldsymbol{I_{Ls}}$ and $\boldsymbol{I_r}$ are known variables that can be obtained directly via Fourier transform, i.e.,

$$I_{Ls_{<k>}} = \begin{cases} -i0.5I_{Ls}, & if\ k = M_1 \\ i0.5I_{Ls}, & if\ k = -M_1 \\ 0 & else \end{cases} \tag{12}$$

$$I_{r_{<k>}} = \begin{cases} I_{Ls}/\pi, & if\ k = 0 \\ -i0.25I_{Ls}, & if\ k = M_1 \\ i0.25I_{Ls}, & if\ k = -M_1 \\ -I_{Ls}/3\pi & if\ k = 2M_1 \\ -I_{Ls}/3\pi & if\ k = -2M_1 \\ 0 & else \end{cases} \tag{13}$$

Similarly, for $v_{DC}(t)$, $i_L(t)$ and $v_o(t)$, which are unknown variables, they can be expressed as

$$v_{DC}(t) \approx \sum_{k=-2M_1}^{2M_1} v_{DC<k>}e^{i2\pi k f_{base}t} = \boldsymbol{E}\boldsymbol{V_{DC}} \tag{14}$$

$$i_L(t) \approx \sum_{k=-2M_1}^{2M_1} I_{L<k>}e^{i2\pi k f_{base}t} = \boldsymbol{E}\boldsymbol{I_L} \tag{15}$$

$$v_o(t) \approx \sum_{k=-2M_1}^{2M_1} v_{o_{<k>}}e^{i2\pi k f_{base}t} = \boldsymbol{E}\boldsymbol{V_o} \tag{16}$$

where

$$\boldsymbol{V_{DC}} = [V_{DC_{<2M_1>}},\dots,V_{DC_{<2>}},V_{DC_{<1>}},V_{DC_{<0>}},V_{DC_{<-1>}},V_{DC_{<-2>}},\dots,V_{DC_{<-2M_1>}}]^T \tag{17}$$

$$\boldsymbol{I_L} = [I_{L_{<2M_1>}},\dots,I_{L_{<2>}},I_{L_{<1>}},I_{L_{<0>}},I_{L_{<-1>}},I_{L_{<-2>}},\dots,I_{L_{<-2M_1>}}]^T \tag{18}$$

and

$$\boldsymbol{V_o} = [V_{o_{<2M_1>}},\dots,V_{o_{<2>}},V_{o_{<1>}},V_{o_{<0>}},V_{o_{<-1>}},V_{o_{<-2>}},\dots,V_{o_{<-2M_1>}}]^T \tag{19}$$

*2: Multi-frequency-domain convolution.* The switching function $s_{sw}(t)$ can be expanded as

$$s_{sw}(t) \approx \sum_{k=-4M_1}^{4M_1} s_{sw<k>}e^{i2\pi k f_{base}t} \tag{20}$$

in which the Fourier coefficients of $s_{sw}(t)$ are

$$s_{sw<k>} = \begin{cases} D, & if\ k = 0 \\ \frac{M_2}{2k\pi i}\left(1 - e^{-i2\frac{k}{M_2}\pi D}\right), & if\ k = M_2, 2M_2, 3M_2 \dots \\ -\frac{M_2}{2k\pi i}\left(1 - e^{i2\frac{k}{M_2}\pi D}\right), & if\ k = M_2, -2M_2, -3M_2 \dots \\ 0, & else \end{cases} \tag{21}$$

By substituting (15) and (20) into the multiplication term in (4), the multiplications term $s_{sw}(t)i_L(t)$ is simplified as

$$\begin{aligned} s_{sw}(t)i_L(t) &= \left(\sum_{k=-4M_1}^{4M_1} s_{sw_{<k>}}e^{i2\pi k f_{base}t}\right) \times \left(\sum_{k=-2M_1}^{2M_1} I_{L<k>}e^{i2\pi k f_{base}t}\right) \\ &= \sum_{k=-2M_1}^{2M_1}\left[\left(\sum_{n=k-2M_1}^{2M_1+k} s_{sw_{<n>}}I_{L<k-n>}\right)e^{i2\pi k f_{base}t}\right] \end{aligned} \tag{22}$$

Equations (17) can be rearranged as

$$s_{sw}(t)i_L(t) = \boldsymbol{E}\boldsymbol{S_{sw}}\boldsymbol{I_L} \tag{23}$$

where the (4$M_1$+1)×(4$M_1$+1) convolution matrix of $\boldsymbol{S_{sw}}$ is

$$\boldsymbol{S_{sw}} = \begin{bmatrix} s_{sw_{<0>}} & s_{sw_{<1>}} & \cdots & \cdots & \cdots & \cdots & s_{sw_{<4M_1>}} \\ s_{sw_{<-1>}} & s_{sw_{<0>}} & \cdots & \cdots & \cdots & \cdots & s_{sw_{<4M_1-1>}} \\ \vdots & \vdots & \ddots & \vdots & \vdots & \vdots & \vdots \\ \vdots & \cdots & s_{sw_{<-1>}} & s_{sw_{<0>}} & s_{sw_{<1>}} & \cdots & \vdots \\ \vdots & \vdots & \vdots & \vdots & \ddots & \vdots & \vdots \\ s_{sw_{<-4M_1+1>}} & \cdots & \cdots & \cdots & \cdots & s_{sw_{<0>}} & s_{sw_{<1>}} \\ s_{sw_{<-4M_1>}} & \cdots & \cdots & \cdots & \cdots & s_{sw_{<-1>}} & s_{sw_{<0>}} \end{bmatrix} \tag{24}$$

Similarly, by substituting (14) and (20) into the multiplication term in (5), the multiplications term $s_{sw}(t)v_{DC}(t)$ is expressed as by involving the convolution matrix of $\boldsymbol{S_{sw}}$

$$s_{sw}(t)v_{DC}(t) = \boldsymbol{E}\boldsymbol{S_{sw}}\boldsymbol{V_{DC}} \tag{25}$$

Consequently, the matrix of $\boldsymbol{S_{sw}}$ can be used to represent the convolution between different Fourier coefficients.

*3: Multi-frequency-domain differentiation.* By substituting the Fourier expansion of $v_{DC}(t)$ in (14) into the left-hand-side differential term in (4), we have

$$C_{DC}\frac{\partial v_{DC}(t)}{\partial t}$$
$$= C_{DC}\sum_{k=-2M_1}^{2M_1}\left(\frac{\partial v_{DC<k>}}{\partial t}e^{i2\pi k f_{base}t} + v_{DC<k>}\frac{\partial e^{i2\pi k f_{base}t}}{\partial t}\right)$$
$$= C_{DC}\sum_{k=-2M_1}^{2M_1}\left(\frac{\partial v_{DC<k>}}{\partial t} + i2\pi k f_{base} v_{DC<k>}\right)e^{i2\pi k f_{base}t}$$
$$= \boldsymbol{E}C_{DC}\left(\frac{\partial \boldsymbol{V_{DC}}}{\partial t} + \boldsymbol{\Omega}\boldsymbol{V_{DC}}\right) \tag{26}$$

where the diagonal matrix $\boldsymbol{\Omega}$ is

$$\boldsymbol{\Omega} = \mathrm{Diag}[i4M_1\pi f_{base}, \cdots, i2\pi f_{base}, 0, -i2\pi f_{base}, \cdots, -i4M_1\pi f_{base}] \tag{27}$$

Similarly, by including the diagonal matrix $\boldsymbol{\Omega}$, the differentiation of $i_L(t)$ (see Eq. (5)) and $v_o(t)$ (see Eq. (6)) can be rewritten as

$$L\frac{\partial i_L(t)}{\partial t} = \boldsymbol{E}L\left(\frac{\partial \boldsymbol{I_L}}{\partial t} + \boldsymbol{\Omega}\boldsymbol{I_L}\right) \tag{28}$$
$$C_o\frac{\partial v_o(t)}{\partial t} = \boldsymbol{E}C_o\left(\frac{\partial \boldsymbol{V_o}}{\partial t} + \boldsymbol{\Omega}\boldsymbol{V_o}\right) \tag{29}$$

Table I shows the state and manipulated variables of the system on the time-domain and multi-frequency-domain models. The first category of the table is the Fourier coefficients decomposition of $i_{Ls}(t)$, and $i_r(t)$, $v_{DC}(t)$, $i_L(t)$ and $v_o(t)$. The second category is the multi-frequency-domain convolution of the multiplication terms $S_{sw}(t)i_L(t)$ and $S_{sw}(t)v_{DC}(t)$. The third category is the multi-frequency-domain differentiation of the differentiation terms $C_{DC}\frac{\partial v_{DC}(t)}{\partial t}$, $L\frac{\partial i_L(t)}{\partial t}$, and $C_o\frac{\partial v_o(t)}{\partial t}$.

TABLE I.
STATE AND MANIPULATED VARIABLES OF THE SYSTEM IN THE TIME-DOMAIN AND MULTI-FREQUENCY-DOMAIN MODELS

| Categories | Time-domain variables | Multi-frequency-domain variables |
|---|---|---|
| Fourier coefficients decomposition | $i_{Ls}(t)$<br>$i_r(t)$<br>$i_L(t)$<br>$v_{DC}(t)$<br>$v_o(t)$ | $\boldsymbol{I_{Ls}}$<br>$\boldsymbol{I_r}$<br>$\boldsymbol{I_L}$<br>$\boldsymbol{V_{DC}}$<br>$\boldsymbol{V_o}$ |
| Multi-frequency-domain convolution | $s_{sw}(t)i_L(t)$<br>$s_{sw}(t)v_{DC}(t)$ | $\boldsymbol{S_{sw}I_L}$<br>$\boldsymbol{S_{sw}V_{DC}}$ |
| Multi-frequency-domain differentiation | $C_{DC}\frac{\partial v_{DC}(t)}{\partial t}$<br>$L\frac{\partial i_L(t)}{\partial t}$<br>$C_o\frac{\partial v_o(t)}{\partial t}$ | $C_{DC}(\frac{\partial \boldsymbol{V_{DC}}}{\partial t} + \boldsymbol{\Omega V_{DC}})$<br>$L(\frac{\partial \boldsymbol{I_L}}{\partial t} + \boldsymbol{\Omega I_L})$<br>$C_o(\frac{\partial \boldsymbol{V_o}}{\partial t} + \boldsymbol{\Omega V_o})$ |

*4: Development of multi-frequency model.* By adopting the variables stated in Table I, the time-domain model (Eq. (4)-(6)) can be converted to the multi-frequency model, given as

$$\begin{cases} C_{DC}\left(\frac{\partial \boldsymbol{V_{DC}}}{\partial t} + \boldsymbol{\Omega V_{DC}}\right) = \boldsymbol{I_r} - \boldsymbol{S_{sw}I_L} \\ L\left(\frac{\partial \boldsymbol{I_L}}{\partial t} + \boldsymbol{\Omega I_L}\right) = \boldsymbol{S_{sw}V_{DC}} - \boldsymbol{V_o} \\ C_o\left(\frac{\partial \boldsymbol{V_o}}{\partial t} + \boldsymbol{\Omega V_o}\right) = \boldsymbol{I_L} - \frac{\boldsymbol{V_o}}{R} \end{cases}$$

=>

$$\begin{bmatrix}\dot{\boldsymbol{V}}_{\boldsymbol{DC}}\\ \dot{\boldsymbol{I}}_{\boldsymbol{L}}\\ \dot{\boldsymbol{V}}_{\boldsymbol{o}}\end{bmatrix} = \begin{bmatrix} -\boldsymbol{\Omega} & -\frac{\boldsymbol{S_{sw}}}{C_{DC}} & \boldsymbol{0}_{(4M_1+1)\times(4M_1+1)} \\ \frac{\boldsymbol{S_{sw}}}{L} & -\boldsymbol{\Omega} & -\frac{\mathbb{I}_{(4M_1+1)\times(4M_1+1)}}{L} \\ \boldsymbol{0}_{(4M_1+1)\times(4M_1+1)} & \frac{\mathbb{I}_{(4M_1+1)\times(4M_1+1)}}{C_o} & -\boldsymbol{\Omega} - \frac{\mathbb{I}_{(4M_1+1)\times(4M_1+1)}}{RC_o} \end{bmatrix}\begin{bmatrix}\boldsymbol{V_{DC}}\\ \boldsymbol{I_L}\\ \boldsymbol{V_o}\end{bmatrix} + \begin{bmatrix}\frac{\boldsymbol{I_r}}{C_{DC}}\\ \boldsymbol{0}_{(4M_1+1)\times 1}\\ \boldsymbol{0}_{(4M_1+1)\times 1}\end{bmatrix} \tag{30}$$

where $\mathbb{I}_{(4M_1+1)\times(4M_1+1)}$ is a $(4M_1+1)\times(4M_1+1)$ identity matrix, $\boldsymbol{0}_{(4M_1+1)\times(4M_1+1)}$ is a $(4M_1+1)\times(4M_1+1)$ zero matrix, and $\boldsymbol{0}_{(4M_1+1)\times 1}$ is a $(4M_1+1)\times 1$ zero vector.

In steady state, the left-hand-side differential terms of (30) become zero. Thus, the steady-state results of the multi-frequency model can be written as

$$\begin{bmatrix}\boldsymbol{V_{DC_ss}}\\ \boldsymbol{I_{L_ss}}\\ \boldsymbol{V_{o_ss}}\end{bmatrix} = -\begin{bmatrix} -\boldsymbol{\Omega} & -\frac{\boldsymbol{S_{sw}}}{C_{DC}} & \boldsymbol{0}_{(4M_1+1)\times(4M_1+1)} \\ \frac{\boldsymbol{S_{sw}}}{L} & -\boldsymbol{\Omega} & -\frac{\mathbb{I}_{(4M_1+1)\times(4M_1+1)}}{L} \\ \boldsymbol{0}_{(4M_1+1)\times(4M_1+1)} & \frac{\mathbb{I}_{(4M_1+1)\times(4M_1+1)}}{C_o} & -\boldsymbol{\Omega} - \frac{\mathbb{I}_{(4M_1+1)\times(4M_1+1)}}{RC_o} \end{bmatrix}^{-1}\begin{bmatrix}\frac{\boldsymbol{I_r}}{C_{DC}}\\ \boldsymbol{0}_{(4M_1+1)\times 1}\\ \boldsymbol{0}_{(4M_1+1)\times 1}\end{bmatrix} \tag{31}$$

The solutions obtained from (31) are the Fourier coefficients of $v_{DC}(t)$, $i_L(t)$, and $v_o(t)$ at steady-state operation conditions, which can be used for the beat frequency component analysis.

To verify the accuracy of the multi-frequency model, circuit simulation carried out using PSIM V12.0.3 and the analytical results obtained from (31) are plotted together as shown in Fig. 4−6 for comparison. The parameters of two-stage wireless power receiver system are: $f_1$=200 kHz, $f_2$=185 kHz, $f_b$=15 kHz, $I_{Ls}$ = 1.4 A, $C_{DC}$= 1 $\mu$F, $L$=33 $\mu$H, $C_o$=50 $\mu$F, and $R$=6 Ω.

As illustrated in Fig. 4(a), (b) and (c), the analytical and simulation time-domain waveforms of $v_{DC}(t)$, as well as their corresponding frequency spectrums, are in close agreement. As expected, a beat frequency component (at 15 kHz) is present and its amplitude is higher than those at 185 kHz ($f_1$) and 200 kHz ($f_2$).

Fig. 5(a), (b) and (c) shows that the theoretical model of $i_L$(t) matches the simulation results. The amplitude of the beat frequency component of $i_L(t)$ at 15 kHz is much higher than that at 185 kHz (i.e., switching frequency).

Fig. 6(a), (b) and (c) shows that theoretical model of $v_o(t)$ is in close agreement with the simulation results. It shows that the amplitude of the beat frequency component of $v_o(t)$ at 15 kHz ($f_b$) is much higher than that at 200 kHz. The results given in Fig. 4–6 indicate that the multi-frequency model can accurately describe the multi-frequency components (DC, $f_b$, $f_1$ and $f_2$).

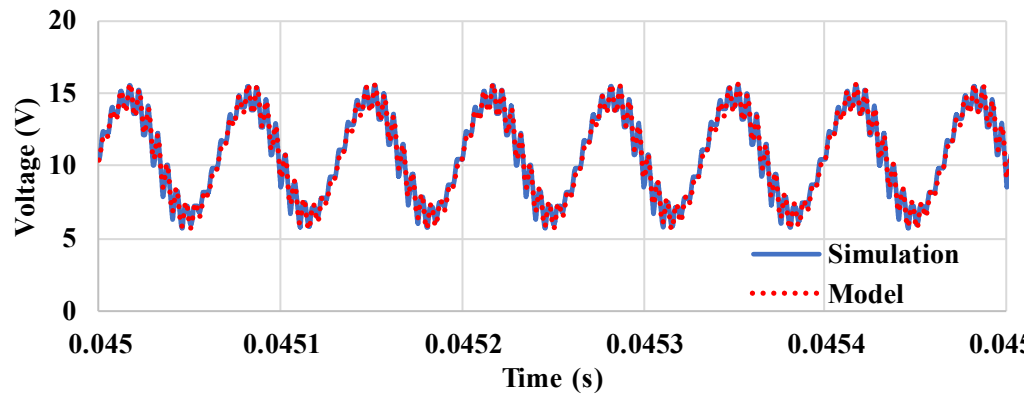

(a) Time-domain waveform (from 0.045 s to 0.0455 s)

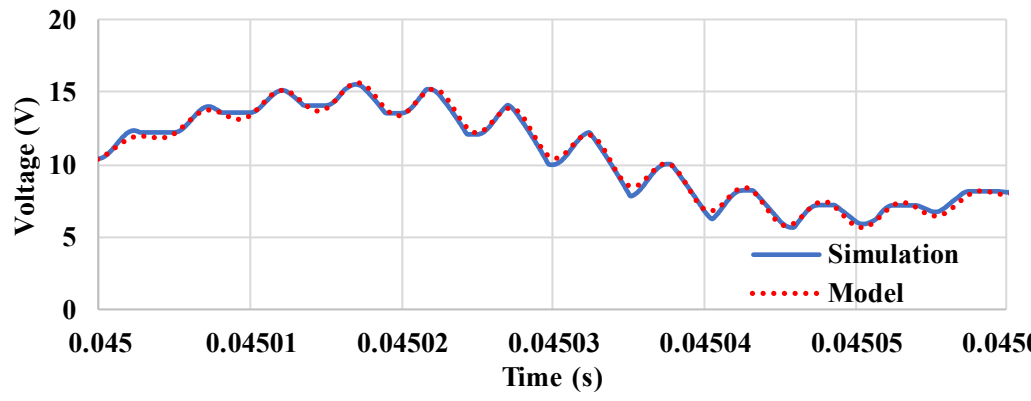

(b) Zoom-in waveform (from 0.045 s to 0.04506 s)

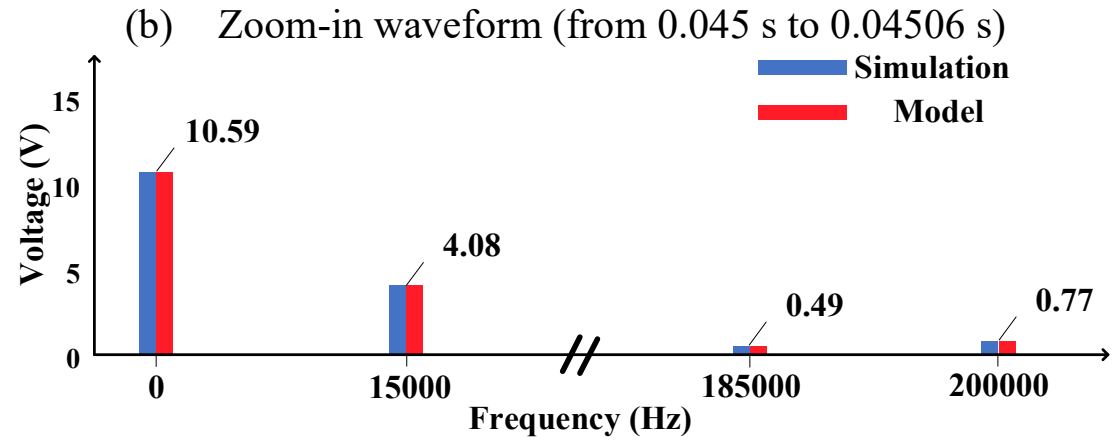

(c) Frequency spectrum

Fig. 4. Simulation and analytical results of $v_{DC}$(t).

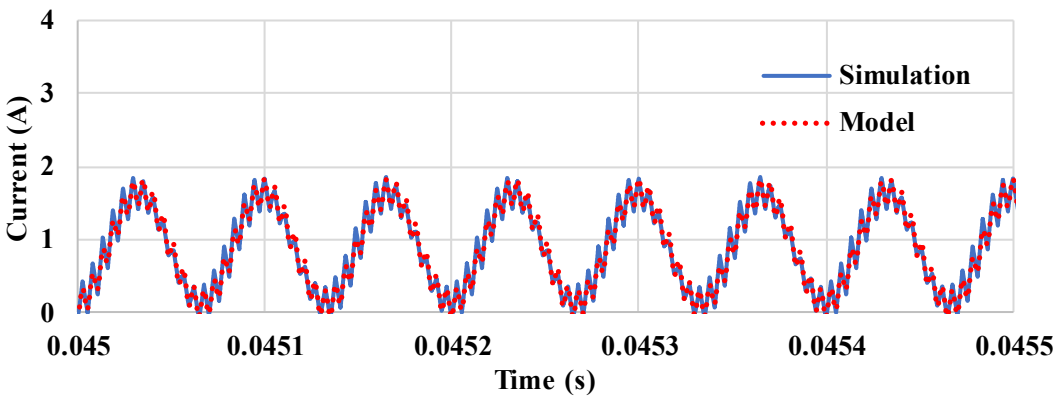

(a) Time-domain waveform (from 0.045 s to 0.0455 s)

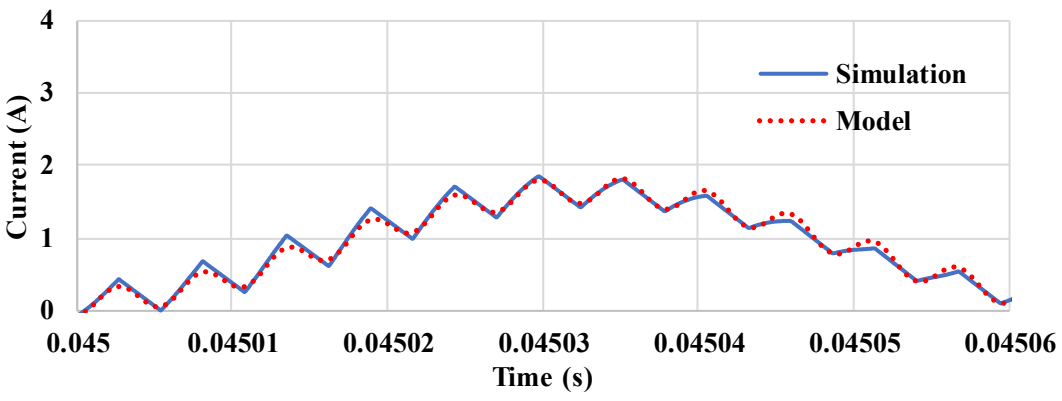

(b) Zoom-in waveform (from 0.045 to 0.04506 s)

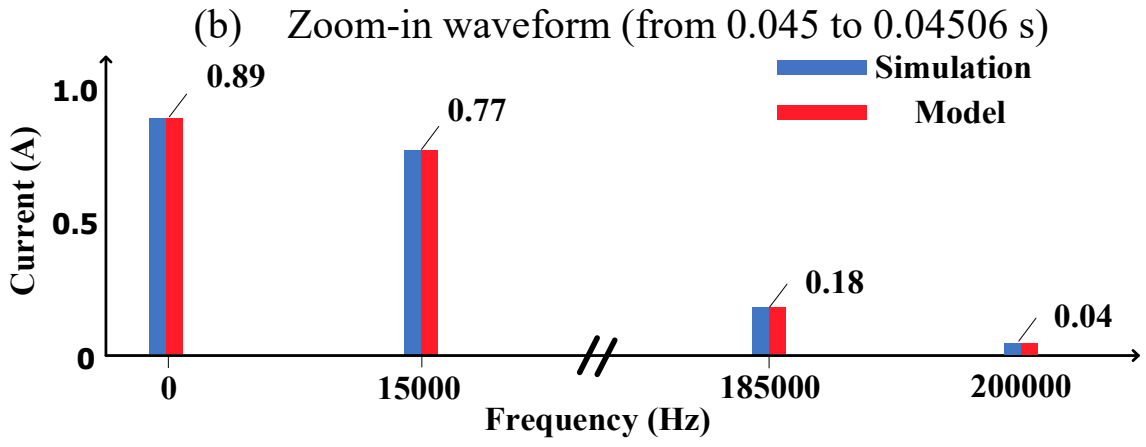

(c) Frequency spectrum

Fig. 5. Simulation and analytical results of $i_L$(t).

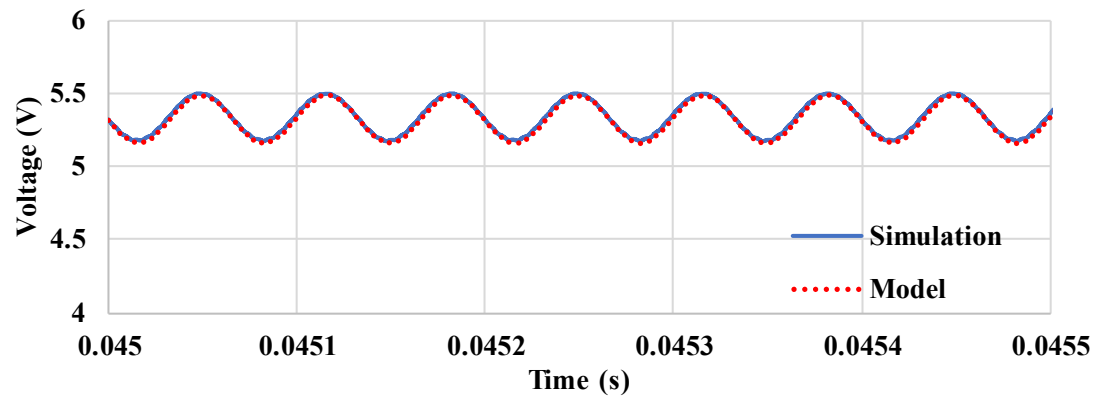

(a) Time-domain waveform (from 0.045 s to 0.0455 s)

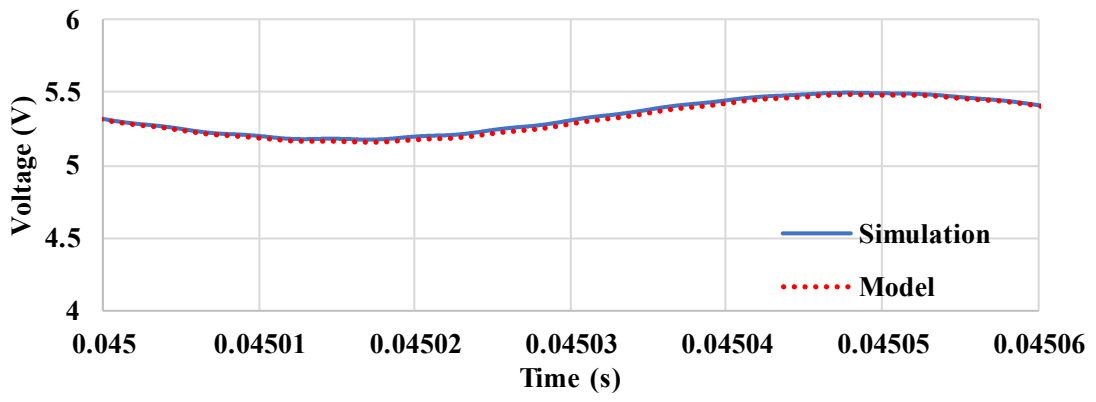

(b) Zoom-in waveform (from 0.045 s to 0.04506 s)

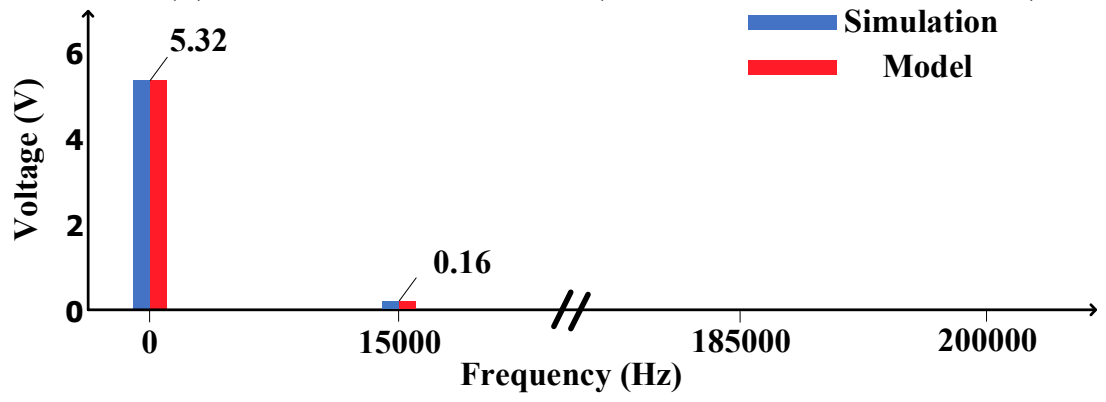

(c) Frequency spectrum

Fig. 6. Simulation and analytical results of $v_o$(t).

### *4. Small-Signal Model and Feedback Control*

The small signal of the system is obtained by linearizing the state-space model Eq. (30), which is written as

$$\begin{bmatrix}\Delta\dot{\boldsymbol{V}}_{\boldsymbol{DC}}\\ \Delta\dot{\boldsymbol{I}}_{\boldsymbol{L}}\\ \Delta\dot{\boldsymbol{V}}_{\boldsymbol{o}}\end{bmatrix}=\begin{bmatrix}-\boldsymbol{\Omega} & -\dfrac{\boldsymbol{S}_{sw}}{C_{DC}} & \boldsymbol{0}_{(4M_1+1)\times(4M_1+1)}\\ \dfrac{\boldsymbol{S}_{sw}}{L} & -\boldsymbol{\Omega} & -\dfrac{\mathbb{I}_{(4M_1+1)\times(4M_1+1)}}{L}\\ \boldsymbol{0}_{(4M_1+1)\times(4M_1+1)} & \dfrac{\mathbb{I}_{(4M_1+1)\times(4M_1+1)}}{C_o} & -\boldsymbol{\Omega}-\dfrac{\mathbb{I}_{(4M_1+1)\times(4M_1+1)}}{RC_o}\end{bmatrix}\begin{bmatrix}\Delta\boldsymbol{V}_{\boldsymbol{DC}}\\ \Delta\boldsymbol{I}_{\boldsymbol{L}}\\ \Delta\boldsymbol{V}_{\boldsymbol{o}}\end{bmatrix}$$
$$+\begin{bmatrix}\boldsymbol{0}_{(4M_1+1)\times(4M_1+1)} & -\dfrac{1}{C_{DC}}\dfrac{d\boldsymbol{S}_{sw}}{dD} & \boldsymbol{0}_{(4M_1+1)\times(4M_1+1)}\\ \dfrac{1}{L}\dfrac{d\boldsymbol{S}_{sw}}{dD} & \boldsymbol{0}_{(4M_1+1)\times(4M_1+1)} & \boldsymbol{0}_{(4M_1+1)\times(4M_1+1)}\\ \boldsymbol{0}_{(4M_1+1)\times(4M_1+1)} & \boldsymbol{0}_{(4M_1+1)\times(4M_1+1)} & \boldsymbol{0}_{(4M_1+1)\times(4M_1+1)}\end{bmatrix}\begin{bmatrix}\boldsymbol{V}_{\boldsymbol{DC_ss}}\\ \boldsymbol{I}_{\boldsymbol{L_ss}}\\ \boldsymbol{V}_{\boldsymbol{o_ss}}\end{bmatrix}\Delta d \tag{32}$$

Fig. 8 shows the simulation data and analytical results of the small-signal response of the DC output voltage $\Delta V_{o<0>}$ to the duty cycle $\Delta d$, where the parameters of the two-stage system remain unchanged. The simulation result and analytical results are quite close, which validates the accuracy of the small-signal model.

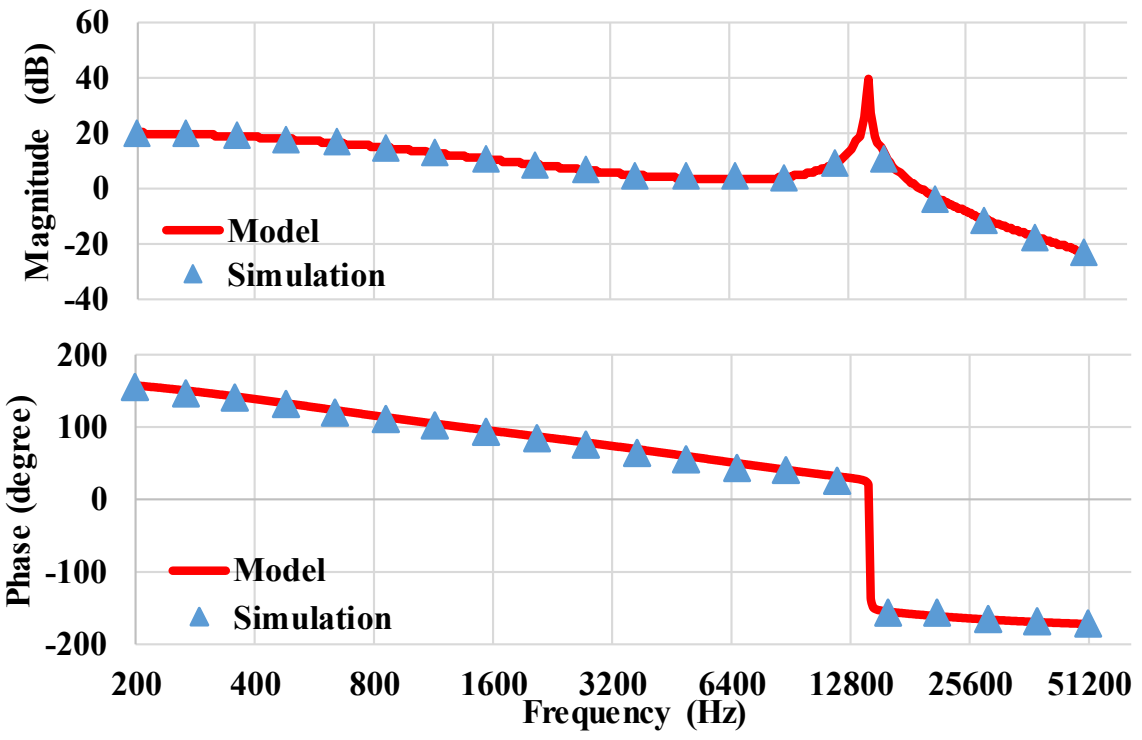

Fig. 8. Bode plots of small-signal response of $\Delta V_{o<0>}$ to $\Delta d$.

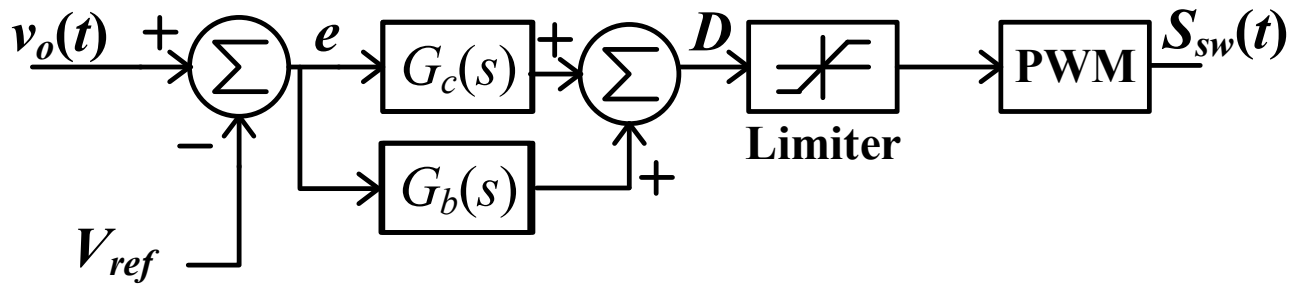

Fig. 9. Overview of the feedback control scheme

To regulate the output voltage, the feedback control scheme (see Fig. 9) is used, where the transfer function of the type II compensator $G_c(s)$ and beat frequency compensator $G_b(s)$ are

$$G_c(s) = k_c \frac{1+\frac{s}{2\pi f_z}}{s\left(1+\frac{s}{2\pi f_p}\right)} \qquad (33)$$
$$G_b(s) = k_b \frac{s}{s^2+(2\pi f_b)^2}$$

The type II compensator is used to minimize the output voltage regulation error, and the beat frequency compensator is used to suppress the beat frequency oscillation [16]. To avoid interference of the beat frequency oscillation, the pole $f_p$ of $G_c(s)$ is designed to be lower than the beat frequency $f_b$. i.e., $f_p<f_b$. The gain $k_c$ and the zero $f_z$ are designed based on the Bode plots of the system [17]. The centre frequency of $G_b(s)$ is located at the beat frequency $f_b$ such that the beat frequency oscillation at the output voltage can be significantly suppressed [16].

By combining the small-signal model Eq. (32) and the transfer function of the compensators Eq. (33), the closed-loop transfer function is obtained. Fig. 10 shows the Bode plots of the open-loop and closed-loop response of the rectifier, where the parameters of the compensators are $k_c$=138.2, $f_z$=100 Hz, $f_p$=10 kHz, $k_b$=700. The closed-loop response presents a loop gain of 47 dB at 1 Hz and 45 dB at 15 kHz, a crossover frequency of 1000 Hz, and phase margin of 96°. The high gain at 1 Hz suggests a low DC regulation error, while the high gain at 15 kHz indicates that beat frequency (15 kHz) oscillation is highly suppressed.

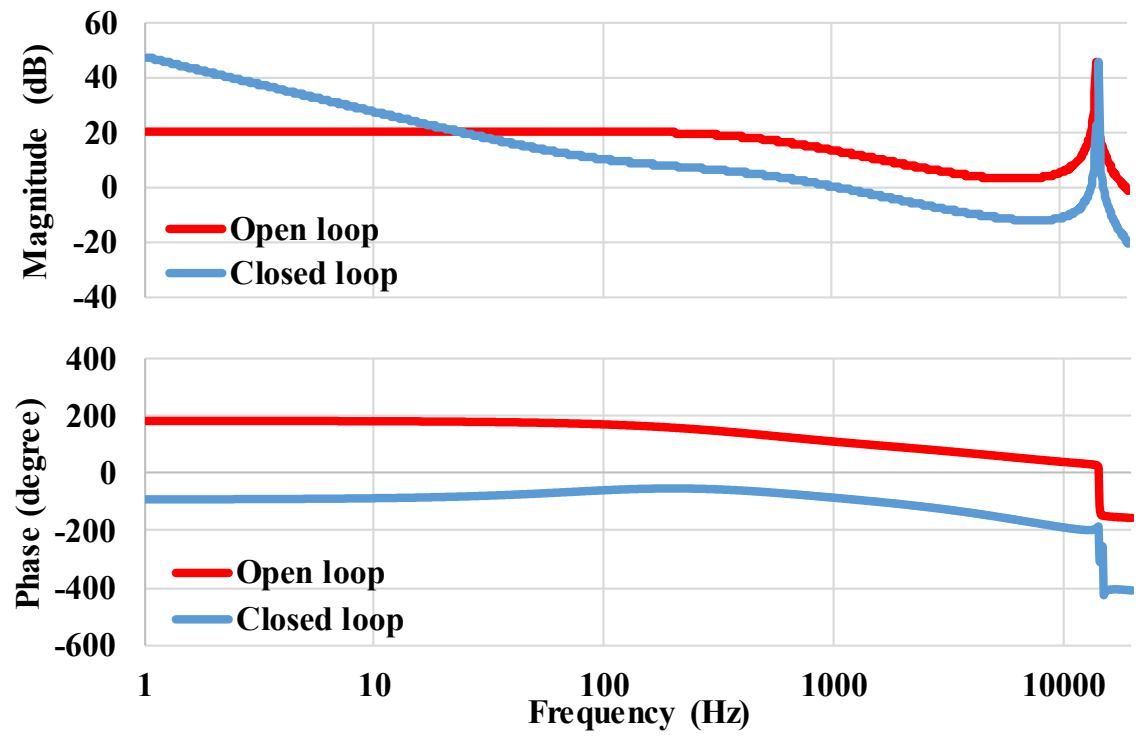

Fig. 10. Bode plots of open loop and closed loop response of the rectifier.

## III. Beat Frequency Oscillation Analysis and Design Solutions

### *A. Beat Frequency Oscillation on the DC-Link and Output Capacitor*

By simplifying Eq. (31), the beat frequency component on the DC-link capacitor $V_{DC<M1-M2>}$ and the output capacitor $V_{o<M1-M2>}$ can be derived as

$$V_{DC<M_1-M_2>} = \frac{Num_{VDC}}{Den} \qquad (34)$$
$$V_{o<M_1-M_2>} = \frac{Num_{Vo}}{Den} \qquad (35)$$

where

$$Num_{VDC} = \frac{D\, I_{\mathrm{Ls}}\, e^{\pi D 2i}\,(1+\pi\, C_o\, R\, f_b\, 2i)\,(e^{\pi D 2i}-1)\, i}{4\, C_{\mathrm{DC}}} \qquad (36)$$

$$Num_{Vo} = \pi\, I_{\mathrm{Ls}}\, R\, f_b\, e^{\pi D 2i}\,(e^{\pi D 2i}-1) \qquad (37)$$

$$\begin{aligned} Den =\ & f_b - 2\, f_b\, e^{\pi D 2i} + f_b\, e^{\pi D 4i} + \pi\, C_o\, R\, {f_b}^2\, 2i - 4\, D^2\, f_1\, \pi^2\, e^{\pi D 2i} \\ & - \pi\, C_o\, R\, {f_b}^2\, e^{\pi D 2i}\, 4i + \pi\, C_o\, R\, {f_b}^2\, e^{\pi D 4i}\, 2i \\ & + 16\, C_{\mathrm{DC}}\, L\, f_1\, {f_b}^2\, \pi^4\, e^{\pi D 2i} \\ & - C_{\mathrm{DC}}\, R\, f_1\, f_b\, \pi^3\, e^{\pi D 2i}\, 8i \\ & - C_o\, D^2\, R\, f_1\, f_b\, \pi^3\, e^{\pi D 2i}\, 8i \\ & + C_{\mathrm{DC}}\, C_o\, L\, R\, f_1\, {f_b}^3\, \pi^5\, e^{\pi D 2i}\, 32i \end{aligned} \qquad (38)$$

Since the beat frequency components $V_{DC<M1-M2>}$ and $V_{o<M1-M2>}$ share the same denominator, the roots of the denominator *Den* will lead to a significant beat frequency oscillation. As a result, the critical frequency $f_{cr}$ is obtained as

$$f_{cr} = root(Den) \qquad (39)$$

### *B. Analysis of the Voltage Oscillation on the DC-link Capacitor*

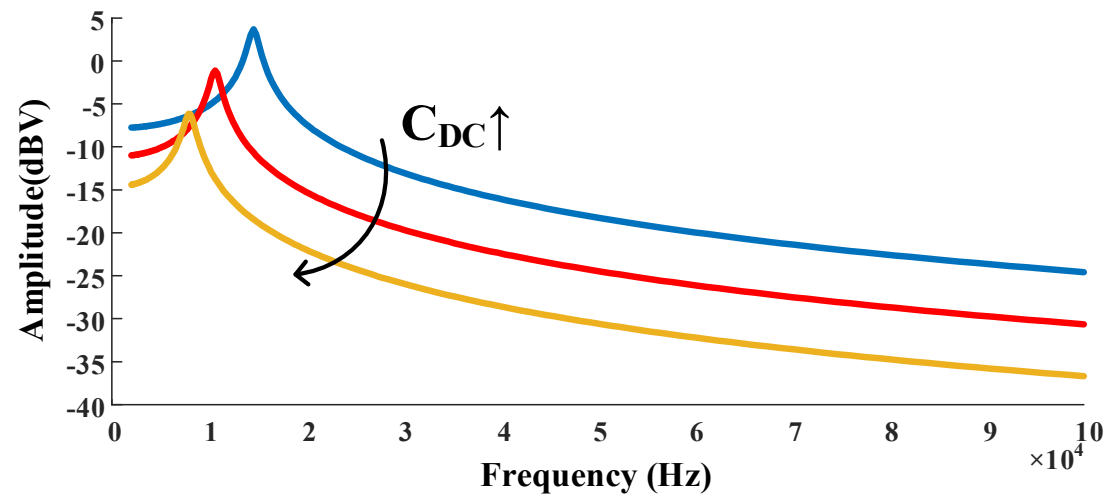

(a) Different $C_{DC}$

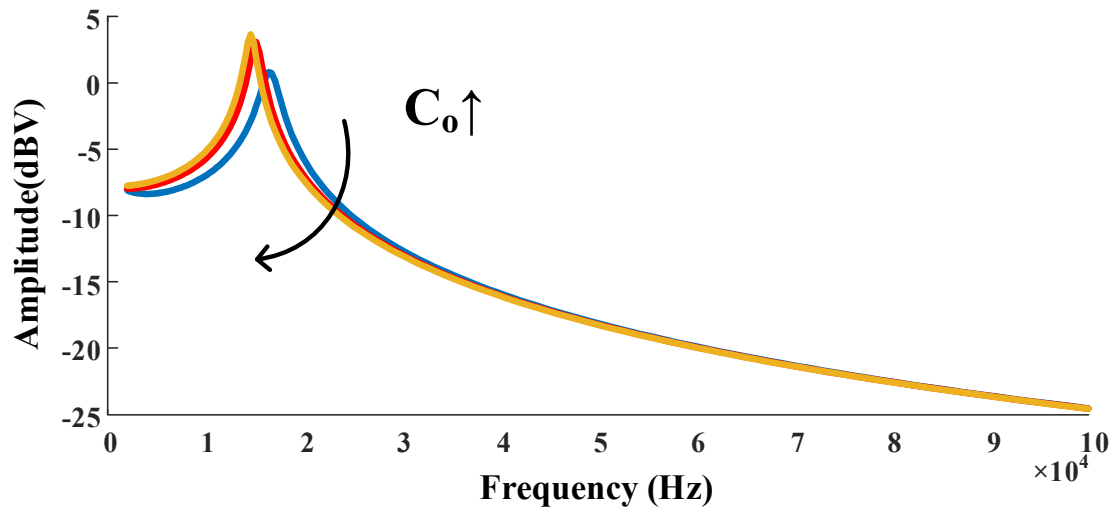

(b) Different $C_o$

Fig. 11. Plots of the normalized beat frequency oscillation $V_{DC<M1-M2>}$ versus beat frequency at different $C_{DC}$ and $C_o$.

Fig. 11 shows the plots of the beat frequency component on the DC-link capacitor versus the beat frequency $f_b$ at the DC-link capacitor $C_{DC}$ and output capacitor $C_o$. Fig. 11(a) shows that the beat frequency oscillation is reduced as $C_o$ increases if $f_b$ is higher than the critical frequency. Fig. 11(b) shows that the normalized oscillation has a slight reduction as $C_o$ increases, if $f_b$ is higher than the critical frequency. These results suggest that the use of a large reactive component and increase $f_b$, such as a large DC-link capacitor and large frequency difference, can effectively limit the beat frequency oscillation of $v_{DC}(t)$.

### *C. Analysis of the Voltage Oscillation on the Output Capacitor*

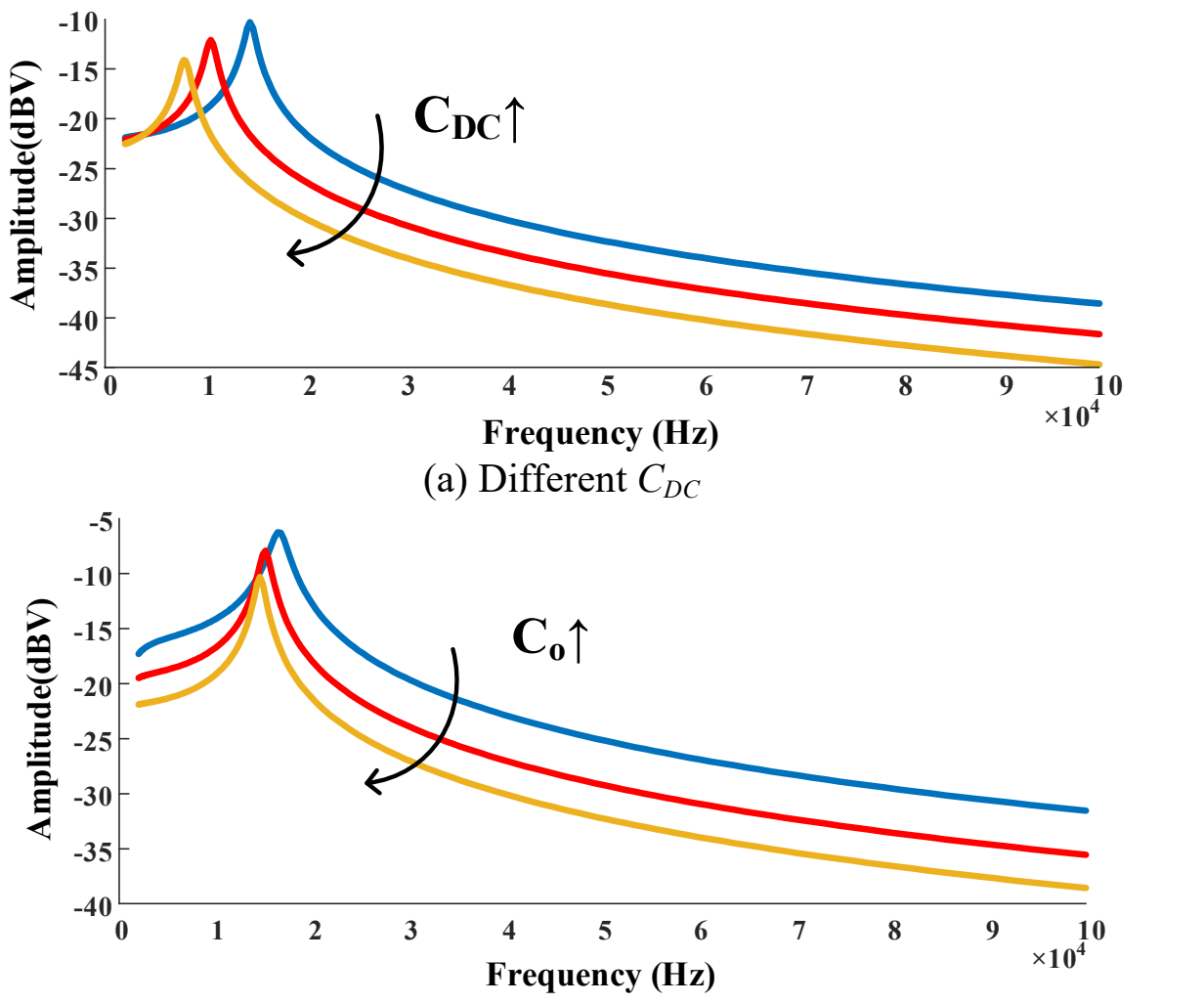

(a) Different $C_{DC}$

(b) Different $C_o$

Fig. 12. Plots of the normalized beat frequency oscillation $\mu_{Co<M_1-M_2>}$ versus beat frequency at different $C_{DC}$ and $C_o$.

Similarly, Fig. 12 shows the plots of the normalized oscillation at beat frequency $\mu_{Co<M_1-M_2>}$ on DC-link capacitor versus beat frequency $f_b$ at different values of $C_{DC}$ and $C_o$. The figure shows that the oscillation is decreased as $C_{DC}$ or $C_o$ increases, as long as $f_b$ is sufficiently high.

Additionally, both Fig. 11 and 12 also show that as $C_{DC}$ or $C_o$ increases, the critical frequency is reduced. Therefore, it is recommended to use large reactive components as well as to broaden the frequency difference in between them to at least 5 times.

### C. Possible Design Solutions

One possible method of alleviating the beat frequency oscillation is to use very large reactive components, e.g., large $C_{DC}$ and $C_o$. This is achievable by using capacitor values of

$$C_{DC} > \max\left\{\frac{I_{Ls}}{2x_{DC}V_{DC<0>}\pi f_1}, \frac{D}{2x_{DC}Rf_2}\right\} \tag{40}$$

$$C_o > \frac{1}{4\pi^2 {f_b}^2 L} \tag{41}$$

where $x_{DC}$ is represents the percentage of amplitude of the beat frequency oscillation to its DC value. The derivation is shown in the appendix. However, this comes at the expense of lower power density and higher cost.

A second possible method is to increase the frequency difference between $f_1$ and $f_2$, such that $f_1 >> f_2$ or $f_1 << f_2$ to avoid the critical frequency. By increasing $f_b$, the beat frequency component is shifted to high frequency such that the beat frequency component can be more effectively absorbed by the capacitors. By considering the distribution of the critical frequency, $f_1$ and $f_2$ can be chosen as

$$f_1 > 5f_2 \text{ or} \tag{42}$$

$$f_2 > 5f_1 \tag{43}$$

Typically, the resonant frequency $f_1$ is pre-determined by wireless power transmitter design. If equation (42) is applied, the buck converter has to operate at a low frequency, which leads to a lower power density. If equation (43) is applied, the buck converter has to operate at a relatively high switching frequency, which has a higher power loss and electromagnetic interference (EMI) issues.

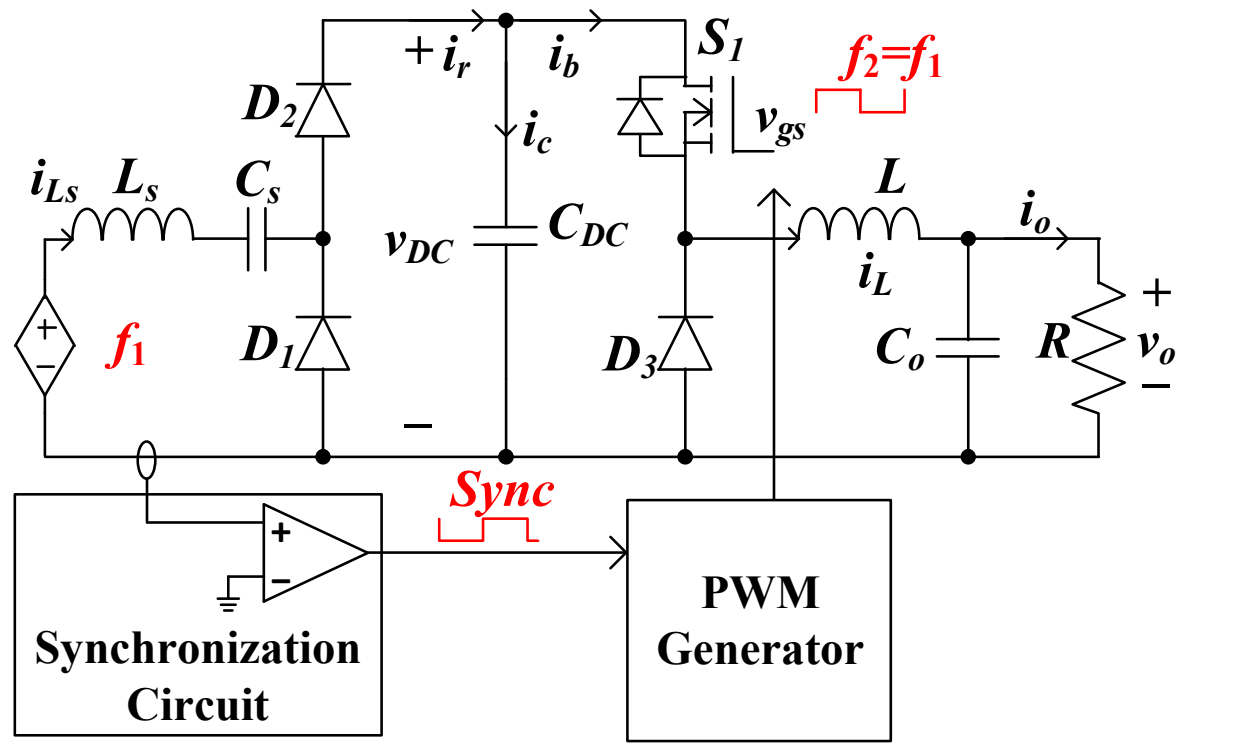

Fig. 13. Two-stage wireless power receiver using frequency-synchronized buck converter.

A third possible method is to synchronize the switching frequency $f_2$ of the buck converter to match the resonant frequency $f_1$, such that $f_2$=$f_1$ and $f_b$=0. By using such a frequency-synchronized buck converter, the interaction between the two power stages that causes the beat frequency oscillation, is avoided. Fig. 13 shows the implementation of such a solution, which contains the two-stage wireless power receiver, an external synchronization circuit, and a PWM generator. The external synchronization circuit senses and converts the sinusoidal input $i_{Ls}$ into a square wave *Sync*. The frequency of *Sync* is $f_1$ and its sharp rising and falling edges are used to trigger the synchronization module of the PWM generator. After receiving *Sync*, the PWM generator generates frequency synchronized PWM signals for the buck converter, where $f_1 = f_2$ or $f_b$=0. This solution is relatively inexpensive.

## IV. Experimental Verification

The solution of synchronizing frequency of the buck converter to match the resonant frequency is chosen for experimental verification. The prototype of the two-stage wireless power receiver (see Fig. 1) and the solution (see Fig. 13) are constructed for comparative study. The parameters and components of the prototype are shown in Table I. Fig. 14 shows a photograph of the experimental setup, of which DSXO3024T oscilloscope, N2790A voltage probe and 1147B current probe are used for the measurement.

Table I. List of Components

| Part | Value / Part Number |
|---|---|
| $L_s$ | 164 $\mu$H ($d$=29 cm, air core) |
| $C_s$ | 3.86 nF |
| $C_{DC}$ | 1 $\mu$F |
| $L$ | 33 $\mu$H |
| $C_o$ | 50 $\mu$F |
| $R$ | 6 Ω |
| Duty ratio $D$ | 50% |
| Gate Driver | ADuM3223 |
| Synchronization circuit | AS-100 (Current Transformer)<br>LM 393P (Comparator) |
| MOSFET | TK56A12N1 |
| Diodes ($D_1$, $D_2$, $D_3$) | TK56A12N1 |
| PWM generator | TMS320F28335 |

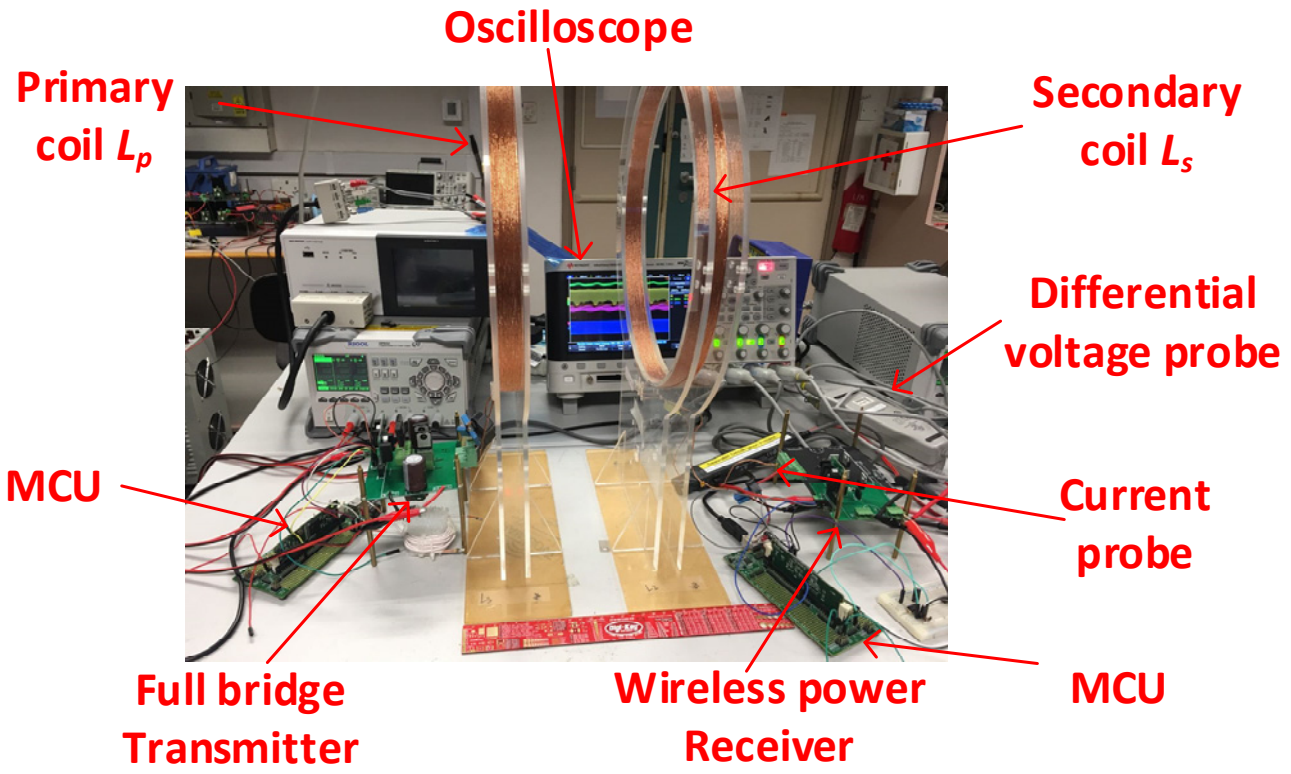

Fig. 14. Photograph of the prototype and the experimental setup.

## A. Time-Domain Performance

Fig. 15 shows the key waveforms of the two-stage wireless power receiver, where $f_1$ and $f_2$ are 200 kHz and 182 kHz, respectively. From Fig. 15(a), it is observed that $v_o$ and $v_{DC}$ contain an oscillation of 18 kHz, which shows the phenomenon of beat frequency oscillation. The peak-to-peak values of $v_o$ and $v_{DC}$ are 0.5 V and 8.38 V, respectively. Fig. 15(b) shows the zoom-in waveforms, in which the peak-to-peak switching ripples of $v_o$ and $v_{DC}$ are 0.25 V and 3.47 V, respectively. Compared with the peak-to-peak values of $v_o$ and $v_{DC}$ shown in Fig. 15(a), the peak-to-peak values of $v_o$ and $v_{DC}$ given in Fig. 15(b) are increased by 100% and 141%. The peak-to-peak value in Fig. 15(a) is contributed by the beat frequency oscillation and switching ripples, while that in Fig. 15(b) is contributed solely by the switching ripple. Therefore, it can be concluded that the increment of peak-to-peak value is mainly attributed to the beat frequency oscillation.

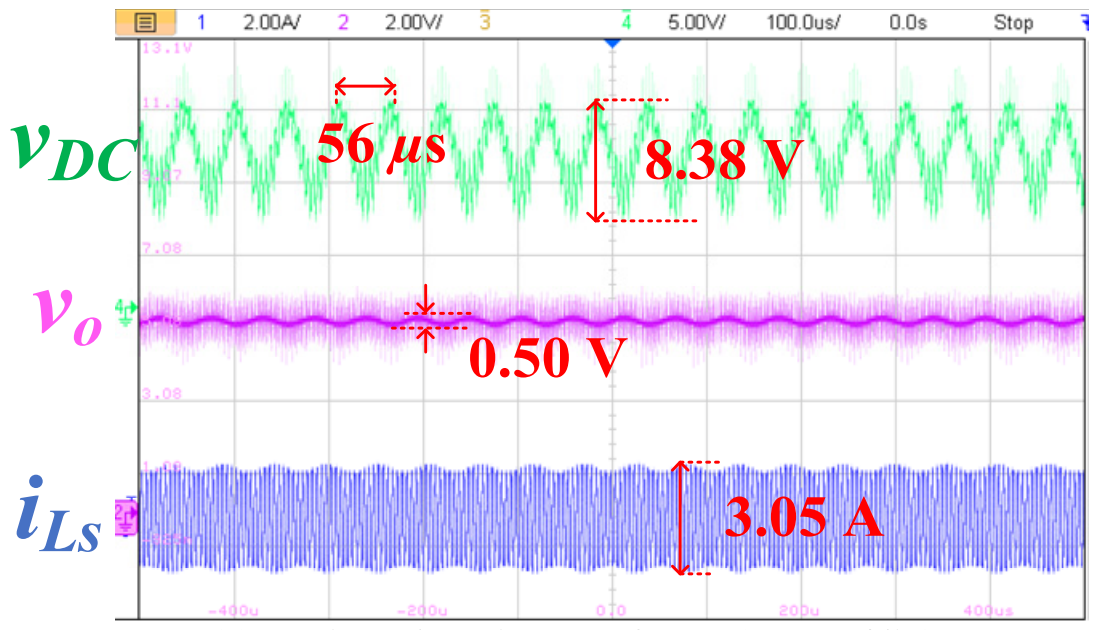

(a) Time-domain waveforms (100 $\mu$s/div)

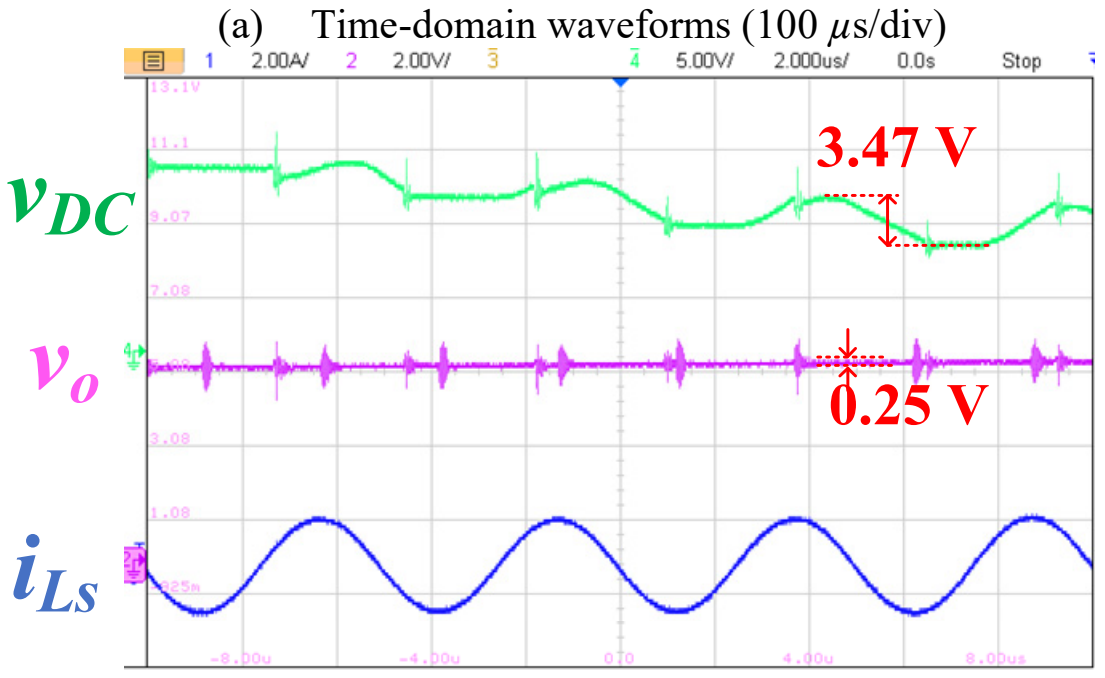

(b) Zoom-in waveforms (2 $\mu$s/div)

Fig. 15. Key waveforms of the two-stage wireless power receiver.

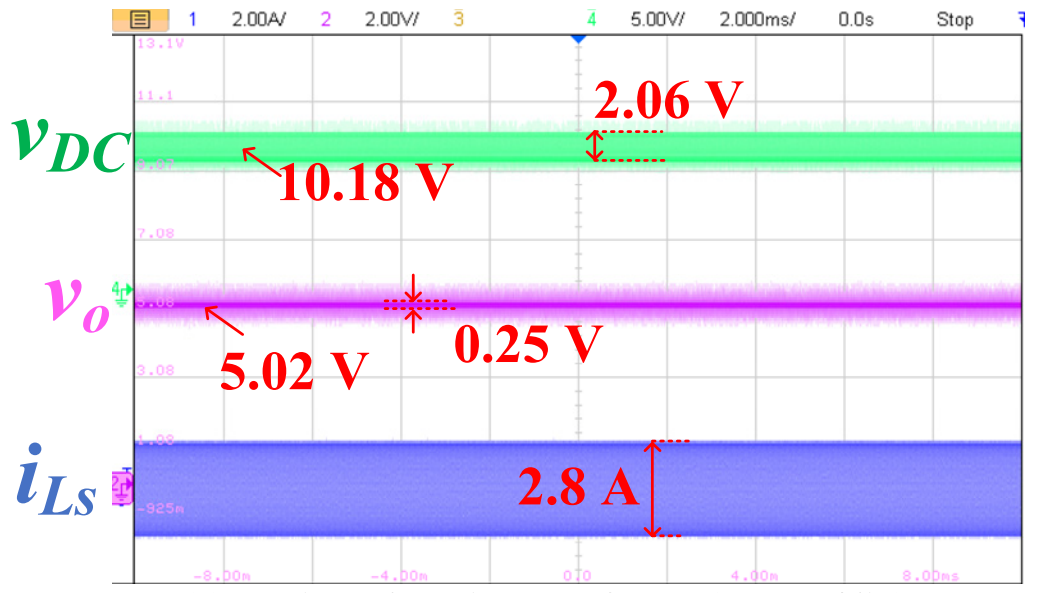

(a) Time-domain waveforms (500 $\mu$s/div)

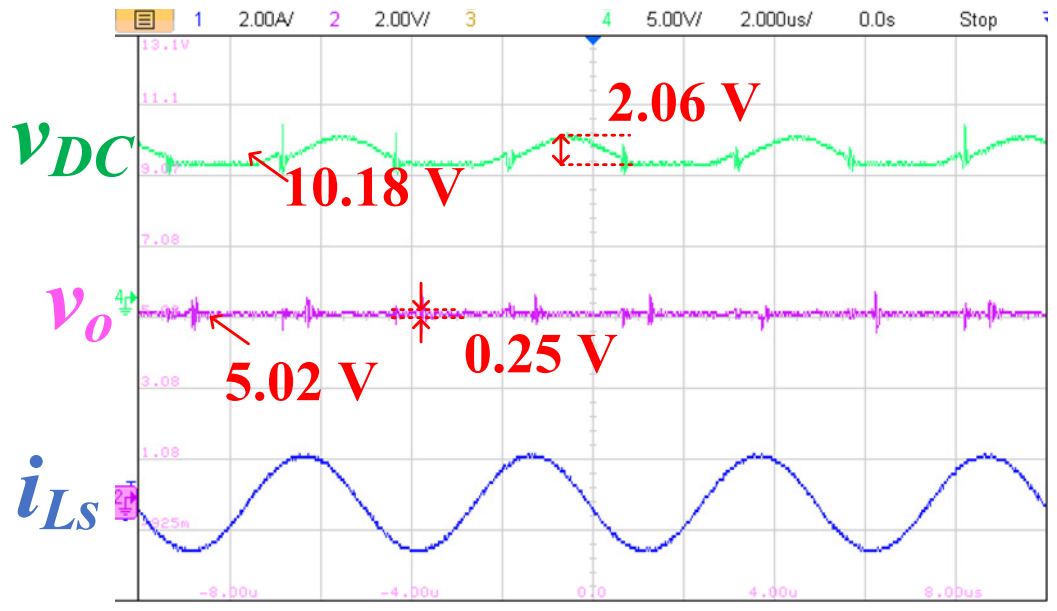

(b) Zoom-in waveforms (2 $\mu$s/div)

Fig. 16. Key waveforms of the two-stage wireless power receiver with frequency-synchronized buck converter.

Fig. 16 shows the key waveforms of the two-stage wireless power receiver using the frequency-synchronized buck converter, where $f_1$ and $f_2$ are 200 kHz. Fig. 16(a) shows that the peak-to-peak value of $v_o$ and $v_{DC}$ are 0.25 V and 2.06 V, respectively. Fig. 16 (b) shows the zoom-in waveforms of $v_o$ and $v_{DC}$, where the peak-to-peak switching ripples of $v_o$ and $v_{DC}$ are 0.25 V and 2.06 V respectively. The peak-to-peak values of the waveforms in different time scales are identical, which suggests that the waveforms in both Fig. 16(a) and (b) contain only the switching ripple and not the beat frequency oscillation. Additionally, as compared to the case without frequency synchronization (waveforms in Fig. 15(a)), the peak-to-peak values of $v_o$ and $v_{DC}$ in this case have been reduced by 50% and 75%, respectively.

## B. Frequency-Domain Performance

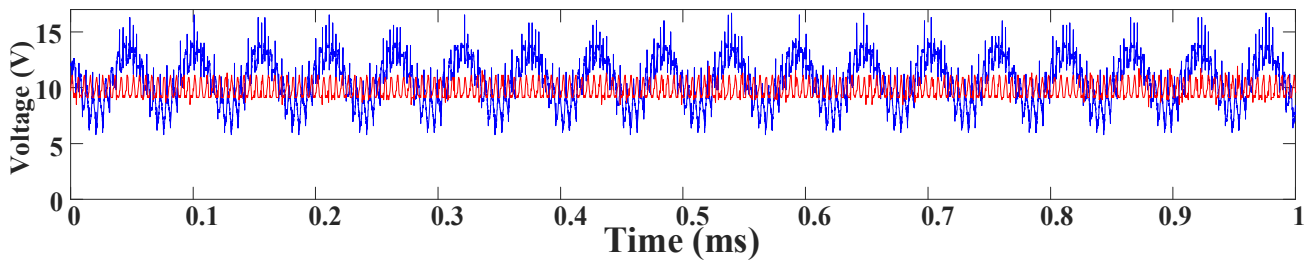

(a) Time-domain waveform of $v_{DC}$

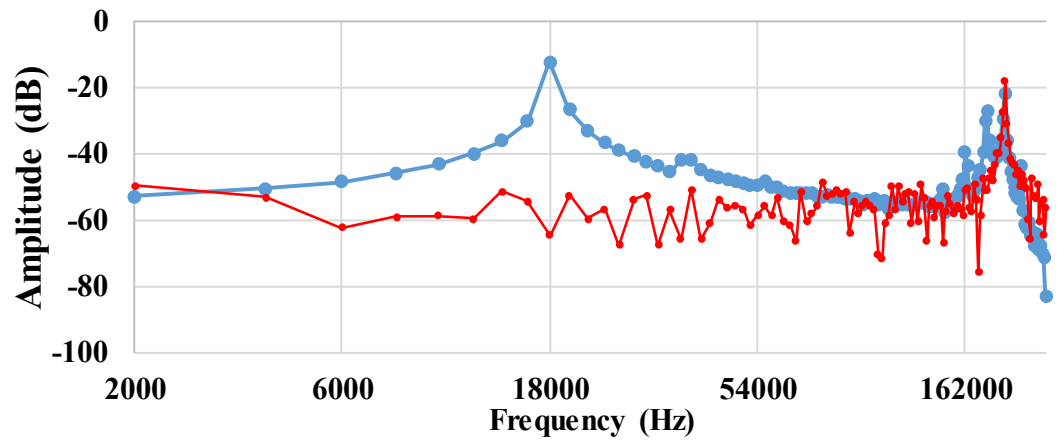

(b) Normalized frequency spectrum of $v_{DC}$

Fig. 17. Time-domain waveform and normalized frequency spectrum of $v_{DC}$ of the two-stage wireless power receiver without frequency synchronization (in blue) and with frequency synchronization (in red).

Fig. 17 shows the time-domain waveform and normalized frequency spectrum of $v_{DC}$ of the two-stage wireless power receiver for the case of not having frequency synchronization (in blue) and that with frequency synchronization (in red). From Fig. 17(a), the DC components are 10.56 V (blue) and 9.93 V (red), respectively. From Fig. 17(b), the normalized spectrum of $v_{DC}$ without frequency synchronization (in blue) shows an amplitude of −12 dB at 1800 Hz, while that with frequency synchronization (in red) has an amplitude of −64 dB at 1800 Hz. This shows that frequency synchronization has reduced the beat frequency component on the DC-link capacitor by 50 dB.

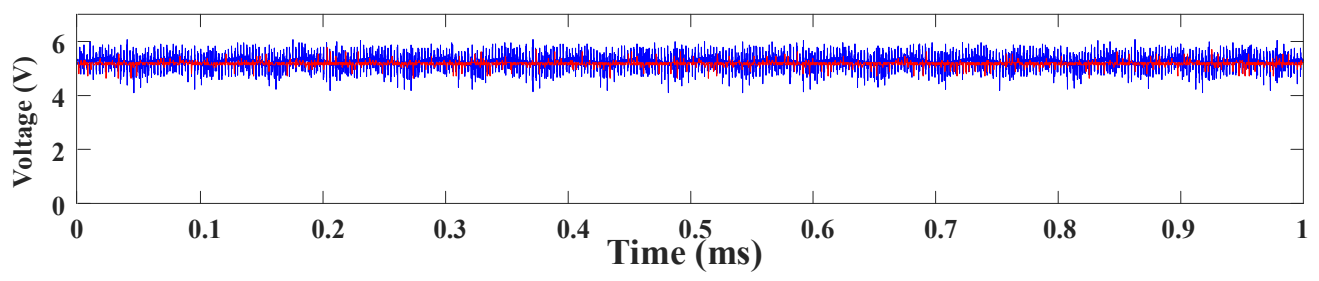

(a) Time-domain waveform of $v_o$

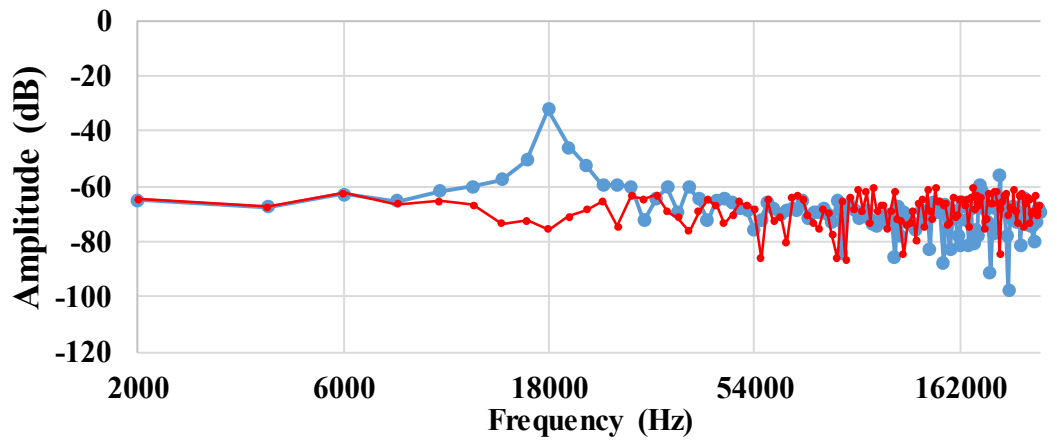

(b) Normalized frequency spectrum of $v_o$

Fig. 18. Time-domain waveform and normalized frequency spectrum of $v_o$ of the two-stage wireless power receiver without frequency synchronization (in blue) and with frequency synchronization (in red).

Fig. 18 shows the corresponding time-domain waveform and normalized frequency spectrum of $v_o$. The DC component of $v_o$ is 5.25 V (blue) and 5.21 V (red), respectively. With frequency synchronization, the voltage ripple is reduced (see Fig. 18(a)). The beat frequency component is suppressed from −32 dB to −75 dB with the use of frequency synchronization (see Fig. 18(b)). Moreover, note that the amplitude of both $v_{DC}$ and $v_o$ over the entire frequency spectrum of 2000 Hz to 250 kHz are lower with frequency synchronization. This shows that not only is the beat frequency oscillation suppressed, but the frequency performance is improved with such a solution. Such an improvement facilitates the use of smaller EMI filters.

Fig. 19 shows the plots of the amplitude of the beat frequency oscillation versus the beat frequency. These curves show that the model and the experimental results are close. As expected, the beat frequency oscillation becomes pronounced around the critical frequency. The small difference at the peak is due to the tolerance and equivalent series resistance of the reactive components. These results suggest that the model is effective and accurate in describing the beat frequency oscillation.

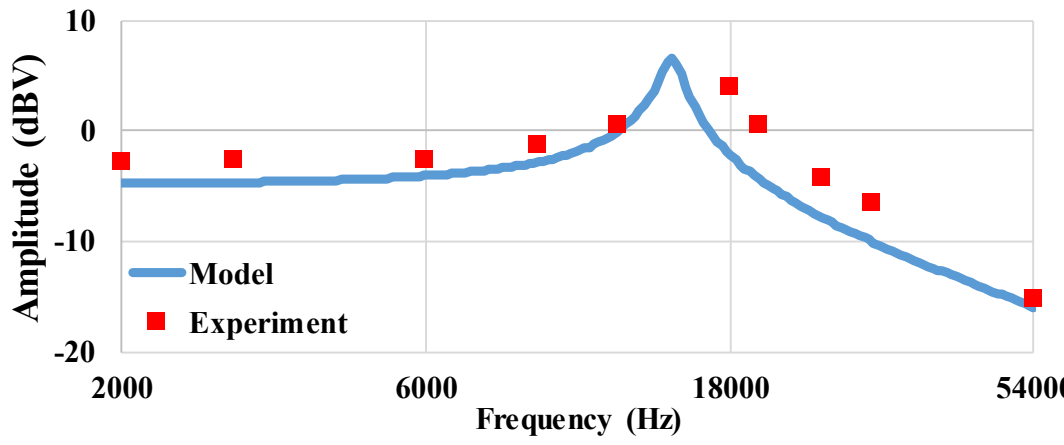

(a) Amplitude of beat frequency oscillation on DC-link capacitor $C_{DC}$

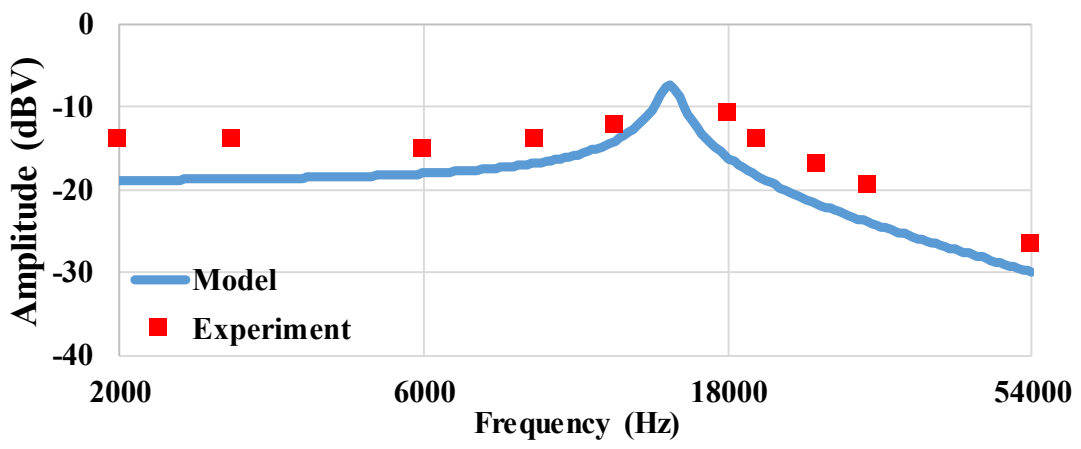

(b) Amplitude of beat frequency oscillation on output capacitor $C_o$

Fig. 19. Plots of the amplitude of beat frequency oscillation on DC-link capacitor and output capacitor versus the beat frequency.

## V. Conclusions

In this paper, the two-stage wireless power receiver system with consideration to the beat frequency oscillation is modelled and analyzed. Various possible solutions on alleviating this oscillation are briefly discussed. Experimental results verifying the presence of beat frequency oscillation in the two-stage wireless power receiver have been provided. Experimental verification on the solution of using a frequency-synchronized buck converter in alleviating the beat frequency oscillation has also been provided. Both time-domain and frequency-domain comparative studies suggest that the solution effectively alleviates the beat frequency oscillation.

## Appendix

### *(A) Design of DC-Link Capacitor*

The DC-link voltage $v_{DC}$ is governed by

$$C_{DC}\frac{\partial v_{DC}(t)}{\partial t} = i_r(t) - s_{sw}(t)i_L(t) \tag{A1}$$

In the analysis, two extremes cases are considered.

The first extreme case happen if $\frac{n}{f_1} < t \le \frac{n+0.5}{f_1}, n \in \mathbb{Z}$ and $\frac{m+D}{f_2} < t \le \frac{m+1}{f_2}, m \in \mathbb{Z}$. Eq. (A1) is simplified as

$$C_{DC}\frac{\partial v_{DC}(t)}{\partial t} = I_{Ls}\sin(2\pi f_1 t) \tag{A2}$$

During this period the ripples can be derived as

$$2\Delta v_{DC}(t) = \frac{\int_{\frac{n}{f_1}}^{\frac{n+0.5}{f_1}} I_{Ls}\sin(2\pi f_1 t)\,dt}{C_{DC}} < x_{DC}V_{DC<0>} \tag{A3}$$

And finally, the DC-link capacitor is designed as

$$C_{DC} > \frac{I_{Ls}}{2x_{DC}V_{DC<0>}\pi f_1} \tag{A4}$$

The second extreme case happen if $\frac{n+0.5}{f_1} < t \le \frac{n+1}{f_1}, n \in \mathbb{Z}$ and $\frac{m}{f_2} < t \le \frac{m+D}{f_2}, m \in \mathbb{Z}$. Eq. (A1) is simplified as

$$C_{DC}\frac{\partial v_{DC}(t)}{\partial t} = -s_{sw}(t)i_L(t) \tag{A5}$$

During this period the ripples can be derived as

$$2\Delta v_{DC}(t) = \frac{Di_L(t)}{C_{DC}f_2} < x_{DC}V_{DC<0>} \tag{A6}$$

And finally, the DC-link capacitor is designed as

$$C_{DC} > \frac{D}{2x_{DC}Rf_2} \tag{A7}$$

To summarize these two extreme cases, one may conclude that

$$C_{DC} > \max\left\{\frac{I_{Ls}}{2x_{DC}V_{DC<0>}\pi f_1}, \frac{D}{2x_{DC}Rf_2}\right\} \tag{A8}$$

### *(B) Design of Output Capacitor*

Since the beat frequency oscillation is transferred from the DC-link capacitor to output capacitor. The *LC* low-pass filter of the buck converter can alleviate the oscillation. Consequently, the corner frequency of the low-pass filter should be less than the beat frequency component. i.e.

$$\frac{1}{2\pi\sqrt{LC_o}} < f_b \quad \text{(B1)}$$

Consequently, the output capacitor can be designed as

$$C_o > \frac{1}{4\pi^2 {f_b}^2 L} \quad \text{(B2)}$$